# Coexistence of strongly buckled germanene phases on Al(111)

W. Wang* and R. I. G. Uhrberg

Department of Physics, Chemistry, and Biology, Linköping University, S-581 83 Linköping, Sweden

Email: weiwa49@ifm.liu.se

## Abstract

We report a study of structural and electronic properties of a germanium layer on Al(111) using scanning tunneling microscopy (STM), low energy electron diffraction and core-level photoelectron spectroscopy. Experimental results show that a germanium layer can be formed at a relatively high substrate temperature showing either (3×3) or (√7×√7)R±19.1° reconstructions. First-principles calculations based on density functional theory suggest an atomic model consisting of a strongly buckled (2×2) germanene layer, which is stable in two different orientations on Al(111). Simulated STM of both orientations fit nicely with experimental STM images and the Ge 3d core-level data decomposed into four components is consistent with the suggested model.

## Keywords

## Introduction

The properties of two-dimensional (2D) materials are currently subjected to intense experimental and theoretical studies. The research is focused on many important properties predicted by theory for various conceivable 2D materials. In similarity with graphene, some other materials are also predicted to show a linear electron dispersion near the Fermi level. Other important properties/phenomena that make 2D materials particularly interesting for incorporation in various devices are, magnetism, superconductivity, Rashba type spin-splitting, quantum spin Hall effect, amongst others. Based on the wealth of physical phenomena exhibited by various 2D materials, they are considered as important future materials of high potential for applications in nano-scale electronics and spintronics.

A sub-group of 2D materials is graphene-like structures formed by the group IV atoms Si, Ge and Sn, i.e., silicene, germanene and stanene. However, Si, Ge, and Sn atoms prefer $sp^3$ hybridization, resulting in a buckled honeycomb structure with a mixture of $sp^2$-$sp^3$ character [1-3]. As a result, the spin-orbital coupling is enlarged and the quantum spin Hall effect is stronger than in graphene [3-6]. The formation of 2D sheets of group IV elements is a great experimental challenge. In this paper we address the germanene case by characterizing a layer of Ge formed on Al(111).

Experimental efforts have been made to grow germanene on metallic substrates as well as band gap materials. Bampoulis et al. [7] proposed a germanene layer with very small buckling (0.2 Å) when they made Pt/Ge crystals by depositing and annealing of Pt on Ge(110). In an inverse case, Li et al. [8] chose Pt(111) as a substrate onto which Ge was evaporated at room temperature. This choice of substrate was motivated by a weaker interfacial interaction compared to other metals with adsorbed two-dimensional sheets such as graphene. They reported that Ge formed a ($\sqrt{19}\times\sqrt{19}$) superstructure on the Pt(111) surface. A model based on a distorted, buckled, germanene sheet was suggested and reported to be consistent with scanning tunneling microscopy (STM) data assuming that only 3 out of 18 Ge atoms inside the ($\sqrt{19}\times\sqrt{19}$) unit cell were observed. These three atoms were about 0.6 Å higher than the rest of the Ge atoms. Later, Švec et al. [9] studied a ($\sqrt{19}\times\sqrt{19}$) superstructure of "silicene" on Pt(111). Based on their theoretical calculation, they believed that a $Si_3Pt$ surface alloy was formed that resembles a twisted kagome lattice. By an extension of their interpretation, they suggested that the ($\sqrt{19}\times\sqrt{19}$) superstructure of Ge on Pt(111) in [8] is also a surface alloy composed of $Ge_3Pt$ tetramers. In another study, Au(111) was chosen as a possible substrate for the formation of germanene because alloy formation was believed to be avoided using this substrate. Deposition of one monolayer (ML) of Ge on Au(111) at ≈200 °C resulted in low energy electron diffraction (LEED) data showing some superstructure spots interpreted as diffraction from ($\sqrt{19}\times\sqrt{19}$), ($\sqrt{7}\times\sqrt{7}$) and (5×5) germanene phases [10]. Only the ($\sqrt{7}\times\sqrt{7}$) periodicity was observed by STM, but the resolution was not sufficient to identify an atomic structure directly from the image. Qin et al. [11] presented results of bilayer germanene on Cu(111) at room temperature. Scanning tunneling spectroscopy showed a "V" shaped density of states, which was also observed by Zhang et al. [12], who synthesized germanene on $MoS_2$ at room temperature. Al(111) was chosen as a substrate to deposit germanene by Derivaz et al. [13], with the motivation that it is a simple unreconstructed metal with surface density of states dominated by s-electrons. A monolayer of Ge formed at a "magic" temperature (in a range of 20 °C around 87 °C) was interpreted as a germanene layer. Well-resolved STM images showed a honeycomb arrangement of blobs corresponding to a (3×3) periodicity with respect to Al(111). An optimized model of (2×2) germanene on a (3×3) Al unit cell was presented. Two Ge atoms were located on top of Al atoms 1.21~1.23 Å higher than the other Ge atoms. Ge deposition at temperatures below the "magic" range was reported to show disorder with a blurred (1×1) LEED pattern, while higher temperatures were reported to result in a sharp (1×1) LEED pattern.

The structure of Ge on Al(111) and the model suggested in [13] were subjected to an investigation using total-reflection high-energy positron diffraction (TRHEPD) by Fukaya et al. [14]. In this study, 1 ML of Ge was deposited on Al(111) held at 350 K. The evaporation rate was ≈ 0.018 ML/min. These parameters are close to the ones in [13] and the formation of a (3×3) superstructure was confirmed by reflection high energy diffraction (RHEED). Interestingly, Fukaya *et al.* arrived at a different conclusion about the model for the (3×3) superstructure. From their TRHEPD data, they concluded that only one Ge atom per unit cell is higher than the other ones. They proposed an explanation to the discrepancy between the STM results in [13] and their results by suggesting that the second Ge atom might be displaced by the interaction with the STM tip (external electric field applied during scanning).

In this paper, we present new data on the Ge/Al(111) system which significantly broadens the view on germanene formation. We show that it is possible to grow well-ordered monolayer Ge at temperatures significantly higher than 87 °C. After deposition at a substrate temperature of ~200 °C, sharp LEED patterns were observed for two phases, i.e., a 3×3 phase and a new √7×√7 superstructure. These phases, formed at higher temperature, deviate from the low temperature phases in the sense that the STM images show hexagonal patterns in contrast to the honeycomb pattern reported in [13]. Our experimental data in combination with DFT calculations lead to a new model that can explain the experimental observations in terms of buckled germanene.

## Results and Discussion

Figure 1(a) shows a LEED pattern, which clearly reveals the coexistence of (3×3) and (√7×√7) periodicities. We find that these two reconstructions coexist with different relative intensities depending on the Ge deposition rate. At a higher rate, e.g. ~0.55 ML/min the (√7×√7) spots appear clearly in the LEED pattern. When Ge is deposited at a lower rate, e.g. ~0.37 ML/min, (3×3) spots dominate, see Fig. 1(b).

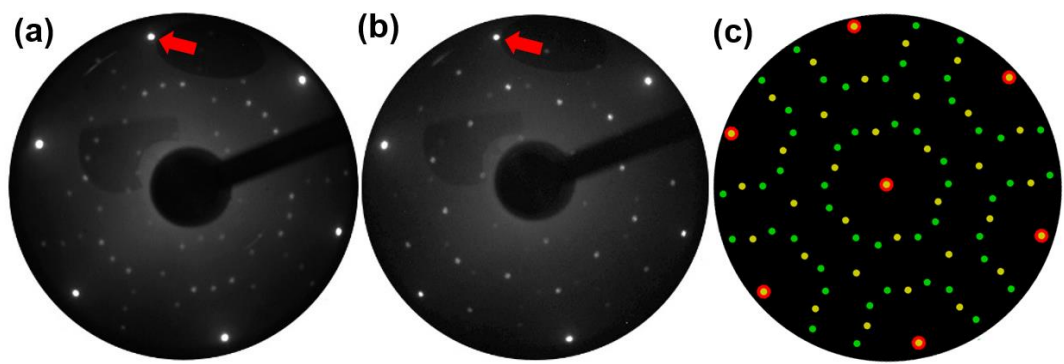

**Figure 1:** (a) LEED pattern obtained at an electron energy of 55 eV from Al(111) with 0.6 ML of Ge deposited at a rate of 0.55 ML/min at a sample temperature of around 200 °C. Diffraction spots corresponding to (3×3) and two domains of (√7×√7) periodicities are clearly observed. (b) LEED pattern obtained at an electron energy of 50 eV from a sample deposited at a rate of 0.37 ML/min. The (3×3) spots are dominating while the (√7×√7) spots are significantly weaker compared to (a). One (1×1) diffraction spot from Al(111) is indicated by a red arrow in (a) and (b). (c) Schematic LEED pattern showing the combination of (3×3) and two domains of (√7×√7) reconstructions. Red circles, yellow spots and blue spots represent Al (1×1), (3×3) and two domains of (√7×√7), respectively.

Considering the theoretical value for the germanene lattice (3.92 ~ 4.06 Å) and the Al(111) surface lattice (2.864 Å) the (2×2) germanene on (√7×√7) Al(111) has less mismatch than on (3×3) Al(111). Furthermore, these two reconstructions have different signs of the stress, so a coexistence could probably reduce the surface energy.

Figure 2(a) is an atomically resolved STM image showing two rotated domains of the (√7×√7) reconstruction with a measured periodicity of ~7.5 Å. The orientations of these two domains are indicated by two blue lines, which are labeled ±19°, respectively. The 0° orientation is indicated by the line in Fig. 2(b), which is an atomically resolved STM image of the (3×3) reconstruction with a measured periodicity of ~8.5 Å. Both reconstructions show a hexagonal structure instead of the honeycomb structure of germanene prepared at low temperature (~87 °C) [13].

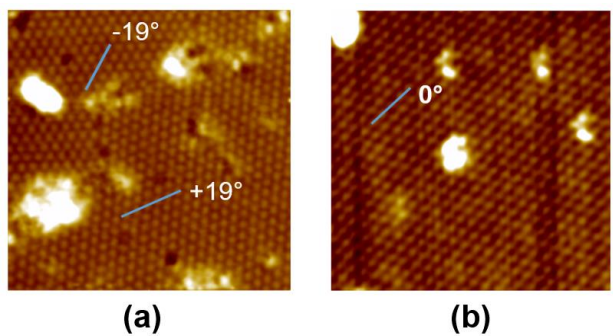

(a) (b)

**Figure 2:** (a) Atomically resolved filled state STM image of a ~19×19 nm$^2$ area showing two rotated hexagonal structures with √7×√7 periodicity. The angle difference between the two blue lines is ~38° which corresponds to the ±19.1° orientations of the two √7×√7 domains with respect to Al(111). (b) Atomically resolved filled state STM image of a ~19×19 nm$^2$ area showing a single hexagonal structure with (3×3) periodicity. Both images were obtained at room temperature with a sample bias of -1.20 V and a tunneling current of 200 pA.

It is interesting to consider the structural results by Fukaya *et al.* [14] obtained from the (3×3) reconstruction prepared in a way similar to that in [13], i.e., at low sample temperature and a low evaporation rate. The results from the TRHEPD technique favored an interpretation of the structure as a germanene layer with one Ge atom per (3×3) cell being higher than the other ones and a corresponding model was presented. However, some restrictions during the relaxation prevented their model from being fully relaxed. Starting from their hexagonal model with one Ge atom higher, we find that it relaxes to the honeycomb structure of the model in [13].

We present a natural modification of the model proposed in [14] that can explain the hexagonal appearance of the Ge layer on Al(111). By making a lateral translation of the germanene layer, one can locate two Ge atoms (Ge(4) and Ge(8) in Fig. 3) above threefold hollow sites of Al(111). Figures 3(a) and 3(c) show fully relaxed atomic models for (2×2) germanene on (3×3)- and (√7×√7)-Al(111). Atom 4 is high while atom 8 is close in height to the rest of the Ge atoms. The height difference (Δz) between Ge(4) and the average level of the other Ge atoms in the (3×3) and (√7×√7) models in Fig. 3 is 2.13 and 1.96 Å, respectively, which is much larger than the values (1.21 ~ 1.23 Å) in [13] and 0.94 Å in [14]. Figures 3(b) and 3(d) show simulated STM images for the (3×3) and (√7×√7) models, respectively, which reproduce the hexagonal structure of the STM images in Fig. 2. The detailed information of these two models is available in Support information.

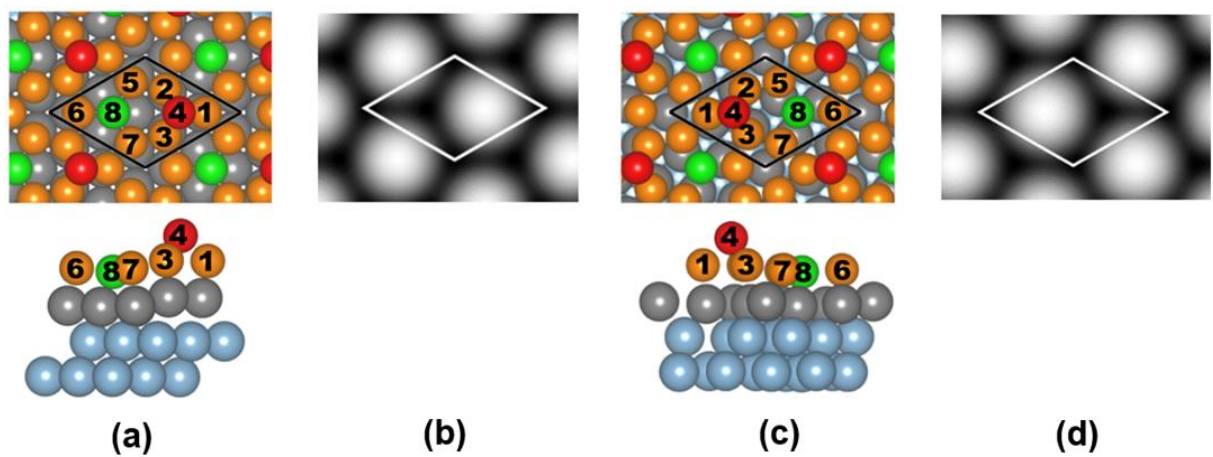

**Figure 3:** (a) Top and side views of the relaxed model of (2×2) germanene on (3×3) Al(111). The black cell represents a (3×3) reconstruction with respect to the Al(111) surface. (b) Calculated STM image generated from the local density of filled states and simulated in a constant current mode at a distance of ~2 Å above Ge(4) for the model in (a). (c) Top and side views of the relaxed model of (2×2) germanene on (√7×√7) Al(111). The black cell represents a (√7×√7) reconstruction with respect to the Al(111) surface. Note that the orientation of the Al(111) substrate is different in (a) and (c). (d) Calculated STM image generated from the local density of filled states and simulated in a constant current mode at a distance of ~2 Å above Ge(4) for the model in (c). The hexagonal structure is consistent with the experimental results in Fig. 2. The highest Ge atom is colored red and labeled Ge(4), the lowest Ge atom is colored green and labeled Ge(8), the other Ge atoms are colored orange, Al atoms are colored light blue, except for the first layer Al atoms which are colored grey. The Ge(4) atom gives rise to the hexagonal pattern observed by STM.

Stephan *et al.* [15] have presented electron spectroscopy data of the Al 2p and Ge 3d core levels for the (3×3) phase. The Ge layer was prepared at a temperature of about 87 °C and a low evaporation rate (0.005 ML/min) as in [13]. Although no structural information was obtained from the sample used for the core-level study, the authors assumed that the surface had the honeycomb type of (3×3) structure reported in [13]. A Ge 3d spectrum, obtained at a photon energy of 130 eV, was fitted by four components of which one only corresponded to 1 or 2 % of the total intensity. The other three components were assigned to three groups of atoms that could be defined from the model. The assignment proposed in [15] implied that the 3d intensity from the two upper Ge atoms was significantly higher than the sum of the intensities from the six remaining Ge atoms. It was suggested that photoelectron diffraction effects could explain the obvious discrepancy between the number of atoms and the core level intensities. An alternative assignment of the 3d components was suggested by Fukaya *et al.* [14]. It was proposed that the two smaller components, one on each side of the main component, should be assigned to the up and down Ge atoms and that the major component corresponds to the six remaining Ge atoms. Unfortunately, the intensities of the different components are not given in [15], so a quantitative evaluation is difficult.

In Fig. 4, we present a Ge 3d core-level spectrum obtained from a surface on which the (3×3) reconstruction was dominating, as verified by LEED patterns at different electron energies. The spectrum was measured using a photon energy of 135 eV in normal emission. A first attempt to analyze the Ge 3d spectrum is based on grouping Ge atoms by their local environment. In this way one can identify four groups of atoms in the (3×3) model, i.e., Ge(1-3), Ge(4), Ge(5-7) and Ge(8). A calculation of the charges was made using the Bader scheme within VASP. The Ge(1-3) atoms gain 0.18-0.19 electron/atom, while Ge(4) loses 0.15 electron. Ge(5-7) atoms gain 0.33-0.34 electron/atom and Ge(8) gains 0.40 electron. Thus from the Bader charge, four distinct groups of Ge atoms can be verified. A fit using four spin-orbit split components is shown in Fig. 4. The relative intensities of components SC1-4 deviate from the expectation that the relative intensities of the four components should be in rough agreement with 1:3:3:1. The intensity of SC2 is quite large while the intensity of SC3 is too small to match an expected intensity of the Ge(5-7) and Ge(1-3) atoms, respectively. Since the LEED pattern showed weak (√7×√7) spots, one has to consider contributions to the Ge 3d spectrum from (√7×√7) domains as well. Based on the Bader charges calculated for the (√7×√7) model, see Support information, one can expect contribution from (√7×√7) mainly to the intensity of SC2 while no intensity is expected at the position of SC3, which provides a plausible explanation to the difference in the SC2 and SC3 intensities. The above discussion of Ge 3d core level spectrum is of course tentative and a rigorous analysis can only be done once truly single-phase samples can be achieved.

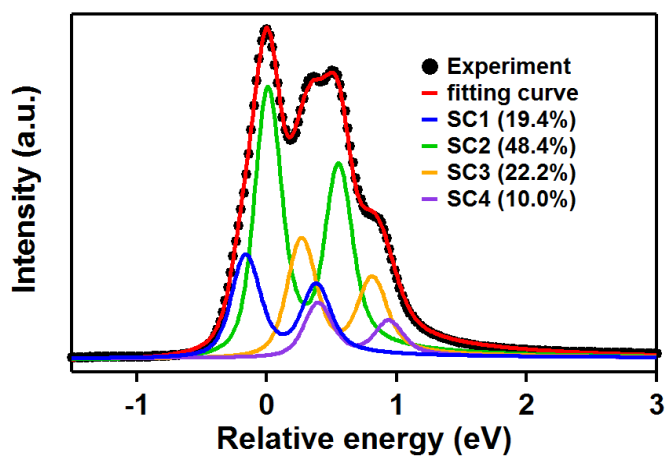

**Figure 4:** Ge 3d core-level spectrum obtained at a photon energy of 135 eV in normal emission. The dots are the experimental data and the fitting curve is the sum of the SC1 to SC4 components. The relative intensities of the four components are shown as a percentage of the total area in the figure. Fitting parameters: Spin-orbit split: 0.545 eV, Branching ratio: 0.67 for SC1 and SC2, 0.63 for SC3 and SC4. Gaussian widths: 177, 163, 183 and 174 meV, respectively, Lorentzian width: 110 meV. The asymmetry parameter of the Doniach–Šunjić line profile is 0.06. The energy shifts with respect to SC2 are -167, +269 and +389 meV.

# Conclusion

We have successfully grown monolayer Ge on Al(111) at a substrate temperature of about 200 °C, which is much higher than the "magic" temperature range mentioned in the literature. Our LEED and STM results confirm a coexistence of two well-ordered hexagonal structures with (3×3) and (√7×√7) periodicities with respect to Al (1×1). Our DFT calculations show that the Ge layer relaxes to a hexagonal structure when two Ge atoms are positioned above threefold hollow sites on Al(111). The experimental and theoretical findings are consistent with a strongly buckled (2×2) germanene layer, which is stable in two different orientations on Al(111). The structural model of the germanene is further supported by simulated STM images. The Ge 3d core-level spectrum can be fitted by four components that are suggested by the calculation of charge.

# Experimental and theoretical details

Samples were prepared *in situ* in two separate ultrahigh vacuum (UHV) systems. One was equipped with LEED and STM (at Linköping University) and the other with LEED and a 2D electron analyzer for photoelectron spectroscopy (at MAX-lab in Lund). A clean Al(111) surface was prepared by repeated cycles of sputtering by $Ar^+$ ions (1 keV) and annealing at approximately 400 °C until a sharp (1×1) LEED pattern was obtained. About 0.6 ML of Ge was deposited at different rates between 0.37 ML/min and 0.55 ML/min while the Al(111) substrate was kept around 200 °C. The reason for depositing less than 1 ML of Ge is to avoid multi-layer formation and the confusion it may lead to. This type of preparation results in a sharp LEED pattern, which is a combination of diffraction from (3×3) and (√7×√7)R±19.1° reconstructions with respect to Al. STM images were recorded at room temperature using an Omicron variable temperature STM in the UHV system at Linköping University. All STM images were measured in constant current mode with a tunneling current of 200 pA. First-principles density functional theory (DFT) calculations were used to investigate the atomic structure of the Ge layer on the Al(111) surface. The structure was modeled by a periodic slab which was built with nine Al layers, one layer of Ge and 15 Å of vacuum spacing. (2×2) germanene was put on top of Al(111)-(3×3) and -(√7×√7)R19.1° supercells, respectively. The positions of all atoms were fully relaxed using the functional of Perdew, Burke and Ernzerhof (PBE) and the projector augmented wave (PAW) method Vienna *ab initio* simulation package (VASP) code [16]. The energy cutoff of the plane-wave basis set was 465 eV, and the k-point mesh was (4×4×1) for both cases. All atoms were relaxed until the average force was within 0.01 eV/Å. Simulated STM images were generated from local density of states according to the Tersoff-Hamann approach [17]. The charge transfer was calculated by the Bader scheme within VASP.

## Acknowledgement

Technical support from Dr. Johan Adell, Dr. Craig Polley and Dr. T. Balasubramanian at MAX-lab is gratefully acknowledged. Financial support was provided by the Swedish Research Council (Contract No. 621-2014-4764) and by the Linköping Linnaeus Initiative for Novel Functional Materials supported by the Swedish Research Council (Contract No. 2008-6582). The calculations were carried out at the National Supercomputer Centre (NSC), supported by the Swedish National Infrastructure for Computing (SNIC).

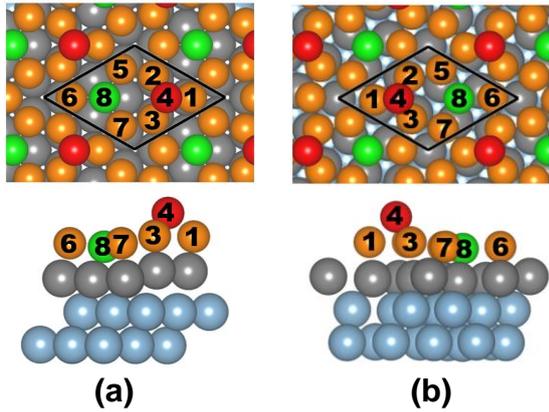

**Figure S1:** (a) Top and side views of the relaxed model of (2×2) germanene on (3×3) Al(111). The black cell represents a (3×3) reconstruction with respect to the Al(111) surface. (b) Top and side views of the relaxed model of (2×2) germanene on (√7×√7) Al(111). The black cell represents a (√7×√7) reconstruction with respect to the Al(111) surface.

In the 3×3 model, the bond lengths and angles between Ge(4) and Ge(1-3) are ~2.56 Å and ~81.0°, respectively; the bond lengths and angles between Ge(8) and Ge(5-7) are ~2.52 Å and ~118.5°, respectively; The adsorption energy is -0.45 eV/Ge atom.

In the √7×√7 model, the bond lengths and angles between Ge(4) and Ge(1-3) are ~2.55 Å and ~78.0°, respectively; the bond lengths and angles between Ge(8) and Ge(5-7) are ~2.55 Å and ~120.0°, respectively; The adsorption energy is -0.46 eV/Ge atom.

The charge transfer is calculated by the Bader scheme within VASP.

The germanene layer gains totally 1.80 and 2.32 electrons in the 3×3 and √7×√7 models, respectively, from Al atoms. In the 3×3 model, Ge(8) gains 0.40 electron; Ge(5-7) gain 0.33-0.34 electron/atom; Ge(1-3) gain 0.18-0.19 electron/atom; Ge(4) loses 0.15 electron. The charge transfer calculation of the √7×√7 model: Ge(5-7) gain 0.38-0.41 electron/atom; Ge(1-3,8) gain 0.31-0.32 electron/atom; Ge(4) loses 0.14 electron.

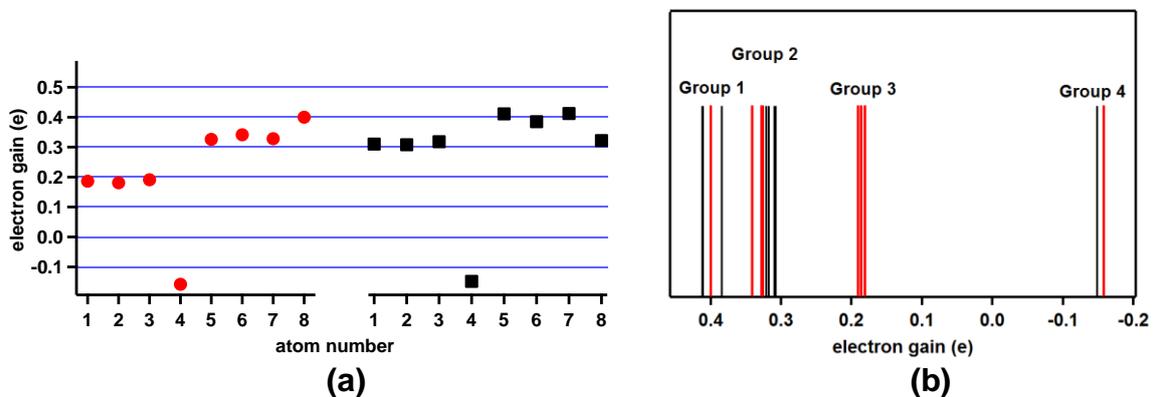

**Figure S2:** (a) Charge transfer diagram for different Ge atoms of the 3×3 (red dots) and the √7×√7 (black squares) models in Fig. S1. (b) The grouping is based on the quantity of electron gain. The red and black bars correspond to the 3×3 and the √7×√7 models respectively. Note that there is no Ge atom of the √7×√7 model with a charge similar to the group 3 atoms of the 3×3 model.